\documentclass[prb,twocolumn,superscriptaddress,showpacs,amsmath,amssymb]{revtex4-1}

\usepackage{graphicx}
\usepackage{latexsym}
\usepackage{amsmath}
\usepackage{amssymb}
\usepackage{amsfonts}
\usepackage{color}
\usepackage{bm}
\usepackage{verbatim}
\usepackage[]{color}
\definecolor{red}{rgb}{1,0,0}
\usepackage{framed}
\definecolor{shadecolor}{RGB}{222,222,221}
\definecolor{MS-color}{RGB}{128,0,128}

\begin{document}

\title{Magnetoelectric effects in superconductor/ferromagnet bilayers}

 \date{\today}
 
\author{D. S. Rabinovich}
\affiliation{Skolkovo Institute of Science and Technology, Skolkovo 143026, Russia}
\affiliation{Moscow Institute of Physics and Technology, Dolgoprudny, 141700 Russia}
\affiliation{Institute of Solid State Physics, Chernogolovka, Moscow
  reg., 142432 Russia}

\author{I. V. Bobkova}
\affiliation{Institute of Solid State Physics, Chernogolovka, Moscow
  reg., 142432 Russia}
\affiliation{Moscow Institute of Physics and Technology, Dolgoprudny, 141700 Russia}

\author{A. M. Bobkov}
\affiliation{Institute of Solid State Physics, Chernogolovka, Moscow reg., 142432 Russia}

\author{M.A.~Silaev}
 \affiliation{Department of
Physics and Nanoscience Center, University of Jyv\"askyl\"a, P.O.
Box 35 (YFL), FI-40014 University of Jyv\"askyl\"a, Finland}

 \begin{abstract}
 We demonstrate that the hybrid structures consisting of
a superconducting layer with an adjacent spin-textured ferromagnet demonstrate the variety of equilibrium  magnetoelectric effects originating from coupling between the conduction electron spin and superconducting current. By deriving and solving the generalized Usadel equation, which takes into account the spin-filtering effect we find that a supercurrent generates spin polarization in the superconducting film which is non-coplanar with the local ferromagnetic moment. The inverse magnetoelectric effect in such structures is shown to result in the spontaneous phase difference across the magnetic topological defects such as a domain wall and helical spin texture. The possibilities to obtain dissipationless spin torques and detect domain wall motion through the superconducting phase difference are discussed. 
 \end{abstract}

  \maketitle

 \section{Introduction}
 
{ Magnetoelectric effects resulting from intrinsic spin-orbit coupling (SOC)  have been studied quite intensively  in different conducting  materials with  
  inversion asymmetry \cite{Levitov1985,Aronov1989,Aronov1991, EDELSTEIN1990233,Kalevich1990,Kato2004,Silov2004,PhysRevLett.86.4358,Ganichev2008}. 
  The direct magnetoelectric effect which is the generation of spin polarization by local electric fields\cite{Levitov1985,Aronov1989,Aronov1991, EDELSTEIN1990233}
  has been experimentally observed by the optical probes\cite{Kalevich1990,Kato2004,Silov2004,Meier2008}, electronic resonance techniques\cite{PhysRevLett.98.187203},  and direct electrical measurements \cite{Li2014,Li2016}.
  Recently, it has become a topic of great interest in connection with magnetic memory applications based on the 
  spin-orbit torque mechanism of magnetization switching\cite{Manchon2008, PhysRevB.79.094422,Gambardella2011,Garello2013,MihaiMiron2010,Chernyshov2009} and domain wall motion\cite{Miron2011}.

  The inverse magnetoelectric effect or the spin galvanic effect is the generation of charge current due to the
   nonequilibrium spin polarization   \cite{Levitov1985}.
  It has been observed experimentally in semiconductors \cite{PhysRevLett.86.4358,Ganichev2008} and normal metals \cite{Sanchez2013}. 

   The  equilibrium counterparts of magnetoelectric effects discussed above exist in superconducting materials 
   resulting from the coupling between supercurrent and various magnetic degrees of freedom. 
   These can be either 
    magnetic moments of conductivity electrons forming spin-triplet Cooper pairs or the localized spins responsible for magnetically ordered states. 
   Up to now the direct magneto-electric effect was reported for superconducting systems 
   in the presence of either  intrinsic\cite{Edelstein1995,Edelstein2005,Malshukov2008,Bobkova2016,Bobkova2017,Bobkova2017_2} 
   or extrinsic SOC \cite{Bergeret2016}. The inverse magnetoelectric effect has been also predicted for systems with intrinsic SOC and under a uniform Zeeman field as in the form of the phase-inhomogeneous superconducting state  \cite{Edelstein1989,Barzykin2002,Samokhin2004,Kaur2005,Dimitrova2007,Houzet2015}, so as in the form of spontaneous electric current \cite{Bobkova2004,Mironov2017,Dolcini2015,Pershoguba2015,Malshukov2016}.
  A number of works predicted an anomalous Josephson effect, which can be viewed \cite{Konschelle2015} as an inverse magnetoelectric effect, specific for Josephson junctions. Its essence is that a spontaneous phase difference intermediate between $0$ and $\pi$ appears in the ground state of the junction. It was proposed for  interlayers with SOC under the applied magnetic field 
   \cite{Krive2004, Asano2007,Reynoso2008, Buzdin2008, Tanaka2009,
 Zazunov2009, Malshukov2010,Brunetti2013,Yokoyama2014,
Bergeret2015,Campagnano2015,Konschelle2015,
Kuzmanovski2016,Zyuzin2016,Silaev2017}
  and  for noncoplanar magnetic interlayers 
  \cite{ Braude2007,   
Grein2009,  Liu2010, 
Kulagina2014,Moor2015,Moor2015a,Bergeret2015,Mironov2015,
Zyuzin2016,Silaev2017,Rabinovich2018}.
 The unified theory
of anomalous Josephson effect which combines the presence of SOC and magnetic texture 
has been developed recently \cite{Bobkova2017a}.
 This effect has been  observed in the Josephson junctions through a quantum dot \cite{Szombati2016} and Bi$_2$Se$_3$ interlayer \cite{Murani2017,1806.01406}.

Given the analogy between intrinsic SOC and the SU(2) gauge field\cite{Tokatly2008} which can be induced by the spatial rotations 
of magnetization in ferromagnets it is natural to expect that 
equilibrium magnetoelectric effect should  exist also in spin-textured 
superconductor/ferromagnet (S/F) hybrid structures. 
 However, in contrast to superconducting systems with SOC, the direct magnetoelectirc effect in 
 S/F hybrids with spin-singlet pairing has not been obtained in the previous theoretical works. 
This is despite that the possibility to produce spin-triplet Cooper pairs in spin-textured S/F systems\cite{Bergeret2001, BergeretRMP2005} and under the simultaneous presence of the homogeneous exchange and SOC \cite{Mel'nikov2012,Bergeret2013,Bergeret2014} is well known and proposed to have a lot of future applications in spintronics and superconducting electronics \cite{Eschrig2015a,Linder2015}.  

The reason for that discrepancy roots in the limitations of the quasiclassical theory, which has been widely used for the study of such systems and which
 neglects the difference in the Fermi velocities and the density of states in spin-up and spin-down subbands. 
Although such approximation is justified for the description of weak ferromagnets like dilute magnetic alloys,
it generically misses the anomalous Josephson effect\cite{Silaev2017} and, as we show below, the direct magnetoelectric effects as well. 

To go beyond the limitations of quasiclassical theory, we consider its minimal extension for a
bilayer S/F system consisting of a thin superconducting film
separated by the tunnel barrier from  the strong spin-textured ferromagnet.
 We demonstrate that this system features both the direct and inverse magnetoelectric effects.
  The latter
produces the phase-inhomogeneous superconducting ground state and can be used for electrical
 detection of the domain wall motion and its chirality. 
The former leads to the generation of non-collinear to the local exchange field spin density in the superconducting film and can be used
for generating the spin torque acting on the spin texture in the adjacent ferromagnet. 

}

 Equilibrium supercurrent-induced spin torques were discussed previously 
in Josephson junctions through the single-domain magnets \cite{Zhu2004,Nussinov2005,Holmqvist2011, Buzdin2008, Konschelle2009},
layered systems \cite{Waintal2002,Linder2011,Halterman2016,Kulagina2014}, in ferromagnetic \citep{Linder2012}
and spin-triplet superconductors \cite{Takashima2017}. Supercurrent-driven domain wall motion in the 
Josephson junction through strong ferromagnet was considered in Ref.\onlinecite{Bobkova2018}.
 
  Here we analyze this effect  for S/F bilayers with singlet superconductors and elucidate its connection with the 
 direct magnetoelectric effect.
 The current-induced spin polarization in a superconducting film is found to have a component perpendicular to 
 the local magnetization in the adjacent ferromagnet.  Therefore, it gives rise to the
 spin  torque acting on the ferromagnet texture $\bm M(\bm r)$. 
 The torque that we have found has two contributions. The first one is similar to the 
 usual adiabatic spin-transfer torque (STT) in normal \cite{Slonczewski1996,Tatara2004,Koyama2011,PhysRevB.70.024417,PhysRevLett.92.207203}
 and superconducting systems\cite{Bobkova2018}.
 The second part of the torque is intimately connected to the local chirality of the magnetization texture and is analogous neither to the adiabatic STT, nor to a non-adiabatic (anti-damping) torque \cite{Zhang2004}.

The paper is organized as follows. In Sec.~\ref{equation} we discuss essential ingredients for obtaining  magnetoelectric effects in S/F bilayers and 
develop a theoretical approach for treating superconductivity in such structures. In Sec.~\ref{phase_shifts} the spontaneous phase gradients in the ground state of the S/F bilayer containing a magnetization texture are obtained and possible applications of this effect for electrical detection of domain walls (DWs) and other magnetic defect presence, motion and chirality are discussed. Sec.~\ref{Sec:torque} is devoted to the calculation and discussion of the supercurrent-induced spin polarization in the superconductor and the resulting torque acting on the DW.

\section{Generalized Usadel equation}
\label{equation}

\begin{figure}[htb!]
 \centerline{\includegraphics[clip=true,width=3.6in]{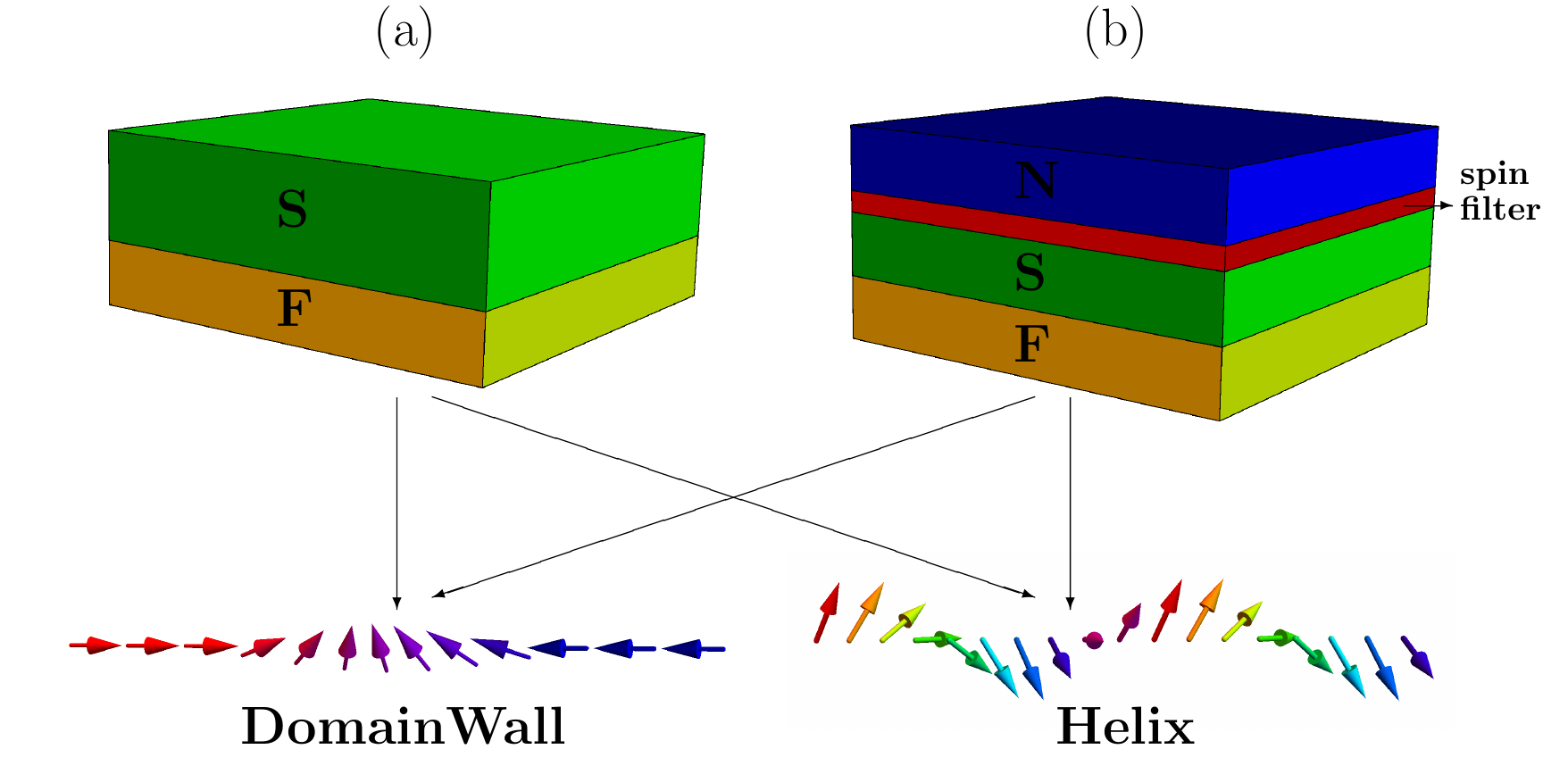}}
 \caption{\label{sketch} 
 Sketch of systems under consideration. The left figure represents a possible realization of S/F bilayer consisting of a conventional superconducting film and a strong ferromagnet. The right  figure is a possible realization of a S/F bilayer with a weak ferromagnet and a spin-filtering interface. The spin filtering interface and the ferromagnet are separated by the superconducting film in order to allow for misaligned magnetizations of these magnetic layers. The considered magnetic texture configurations are shown in the bottom of the figure. 
       }
 \end{figure}

We consider a thin (the thickness $d$ along the $z$ direction is much smaller than the superconducting coherence length) superconducting film in a contact with a ferromagnet. The sketch of different system configurations, which we consider, is presented in Fig.~\ref{sketch}. The  $z$-axis is perpendicular to the film  plane and 
below we denote $\bm r =(x,y)$ to be 2D coordinate vector  in plane of the film.  

It is widely accepted in the literature that if the thickness of the S film $d$ is smaller than the superconducting coherence length $\xi_S$, the magnetic proximity effect, that is the influence of the adjacent ferromagnet with the magnetization $\bm M(\bm r)$ on the S film can be described by adding the effective exchange field $\bm h_{eff}(\bm r) \propto \bm M(\bm r)$ to the quasiclassical Eilenberger or Usadel equation, which is used for treating the superconductor. This was reported as for metallic \cite{Bergeret2001,Bobkova2015}, so as for insulating 
\cite{Tokuyasu1988,PhysRevB.38.4504} ferromagnets. Typically, the resulting effective exchange field is inversely proportional to $d$ and depends on the true exchange field of the ferromagnet, thickness of the ferromagnetic film and interface transparency \cite{Bergeret2001,Bobkova2015,Tokuyasu1988,Cottet2009,Eschrig2015} Effective exchange energy values $h_{eff} \lesssim \Delta$ have been experimentally reported for superconducting films in proximity to the ferromagnetic insulator EuS \cite{Hao1991,
Xiong2011,Wolf2014}. 

In general, the magnetic proximity effect is not reduced to the effective exchange only. The other terms act analogously to additional magnetic impurities in the superconductor \cite{Cottet2009,Eschrig2015} and for the linearized limit considered here can be included into the depairing factor $\Gamma$, see below. 
 
Hence we consider the Usadel equation in the form
 \begin{align}
   -iD \bm \nabla
   (\check g \bm \nabla \check g)+
    \Bigl[ \check \Lambda  - 
    \check \Delta (\bm r),\check g \Bigr] = 0,
    \label{eilenberger_ordinary}
      \end{align}
where 
$\check g \equiv \check g(\bm r, \omega)$ is the momentum-averaged 
quasiclassical Green's function,  $D$ is the diffusion coefficient and $\omega$ is the Matsubara frequency.
The matrix gap function is 
$\check \Delta (\bm r) = \Delta (\bm r) \hat\tau_+ - \Delta^* (\bm r) \hat\tau_-$ with $\hat\tau_{\pm} = (\hat\tau_x \pm i\hat\tau_y)/2$ and the
 diagonal spin-dependent potential term is given by 
  \begin{align}
   \label{Eq:Lambda0}
   & \check\Lambda =\check\Lambda_0 \equiv \hat\tau_z[ i\omega  + 
   {\bm {\hat\sigma}}\bm h_{eff}(\bm r) ].
   \end{align}
  We denote $\hat\sigma_i$ and $\hat\tau_i$ to be the Pauli matrices in spin and particle-hole spaces, respectively.


Although Eq.(\ref{eilenberger_ordinary}) describes the formation of spin-triplet superconducting correlations in the spatially-inhomogeneous field $\bm h_{eff}(\bm r)$,
it completely misses the magnetoelectric effects, which can be understood from the following argument. 
In general, the supercurrent $\bm j$ flowing through the system 
is a function of effective exchange field
$\bm h_{eff} \propto \bm M$ and the superconducting phase gradient $\nabla\varphi (\bm r)$.  
The time reversal symmetry dictates that $\bm j(\nabla\varphi, \bm M) = - \bm j(-\nabla\varphi, -\bm M)$. 
It was shown \cite{Silaev2017} that for a system described by Eq.~(\ref{eilenberger_ordinary}) the additional quasiclassical 
symmetry holds $\bm j(\bm M) = \bm j(-\bm M)$. Combining this with the time-reversal symmetry we obtain 
$\bm j(\nabla\varphi) = -\bm j(-\nabla\varphi)$. Consequently, at $\nabla\varphi=0$ the anomalous supercurrent  and phase-inhomogeneous superconducting ground states are not allowed. To allow for the magnetoelectric effects two conditions should be satisfied simultaneously: (i) the magnetization $\bm M$ of the ferromagnet should be noncoplanar and (ii) the magnetization of the ferromagnet should be treated beyond the quasiclassical approximation Eq.~(\ref{eilenberger_ordinary}) in order to violate the quasiclassical symmetry $\bm j(\bm M)= \bm j(-\bm M)$ \cite{Silaev2017}.

Below we demonstrate that the minimal necessary generalization of the  Usadel equation allowing for describing magnetoelectric effects is to include the "spin-dependent depairing" term which modifies the 
   diagonal potential in Eq.(\ref{eilenberger_ordinary})
  \begin{align}
    \check \Lambda = \check\Lambda_0 + i\Gamma {\rm sgn} \omega  
 (\hat\tau_z + \bm P {\bm{\hat \sigma}} )
 \label{lambda}
  \end{align}
 %
Hence the effective 2D Usadel equation in the superconducting film looks as follows 
\begin{eqnarray}
 - D \partial_{\bm r} \bigl( \check g \partial_{\bm r} \check g \bigr) + \Bigl[ (\omega +  {\rm sgn} \omega \Gamma) \hat\tau_z  -i  \bm h_{eff} 
  \hat {\bm \sigma} \hat\tau_z  +  \nonumber \\
 {\rm sgn} \omega \Gamma \bm P  
 \hat {\bm \sigma} +i\check \Delta_{eff} (\bm r), \check g \Bigr]=0,~~~~~~
\label{usadel_spin_filtering_p}
\end{eqnarray}  
  where 
  $\partial_{\bm r} = (\partial_x, \partial_y) $ and  the effective gap function   $\Delta_{eff}$  can be different from $\Delta$
  due to the presence of the F layer as described below.
  
    Qualitatively the last term in Eq.(\ref{lambda}) describes 
  the suppression of superconductivity in the 
film due to spin-dependent tunneling of electrons forming Cooper pairs
into the adjacent normal ferromagnet and $\Gamma$ is the effective
depairing parameter.
The polarization $\bm P$ describes the efficiency and quantization
axis of the spin-filter which acts on the electrons during the
tunnelling  between the superconductor and
adjacent layer, ferromagnetic or normal. It is just the polarization term $\bm P$ that allows for violating the symmetry $\bm j(\bm M)= \bm j(-\bm M)$. The reason is that in order to obtain $\bm P \neq 0$ one needs to treat the exchange field of the spin-filter layer (a strong metallic ferromagnet or a ferromagnetic insulator) beyond the quasiclassical approximation or, in other words, to take into account the splitting of spin-up and spin-down momenta inside this layer. 

Below by two examples of S/F bilayers with spin-filtering interface we elucidate the physical origin of the spin-filtering term in Eq.(\ref{lambda}). The first example, represented in Fig.~\ref{sketch}(a), is a bilayer system consisting of a superconducting film thinner than the coherence length and a strong ferromagnet. 
The S/vacuum interface at $z=d$ is described by the impenetrable boundary conditions $\check g \partial_z \check g \bigl |_{z=d}=0$. 
The S/F interface is located at $z=0$ and has low transparency. 
This interface can be described by the general boundary conditions containing spin-mixing and spin-filtering terms\cite{PhysRevB.38.4504,Cottet2009,Eschrig2015}. Up to the first order with respect to the spin-mixing angle $\delta \phi_n$, transmission probability $T_n$ and transmission polarization $P_n$ corresponding to the $n$-th transmission channel, these boundary conditions are:
\begin{eqnarray}
\check g \partial_z \check g \bigl |_{z=0}=  [ \hat \Gamma \check g_F \hat \Gamma^\dagger, \check g] - 
i\frac{G_\phi}{2\sigma_S}
[ ({\bm m} {\bm {\hat \sigma}}) \hat \tau_z, \check g],
\label{bc_SF}
\end{eqnarray}
where $\hat \Gamma = u\hat\tau_z + t({\hat {\bm\sigma}} {\bm h})/h$. The spin-dependent tunneling coefficients are determined by  $u^2+ t^2 = G_T/2\sigma_S$ and 
$2ut = G_{MR}/2\sigma_S $ where 
$\sigma_S$ is the normal-state conductivity in the S region, $G_T = G_q \sum_n T_n $ is the junction conductance, $G_{MR} = G_q \sum_n T_n P_n $ accounts for the difference of the junction conductances for spin up and spin down electrons and $G_\phi = G_q \sum_n \delta \phi_n$ is the total spin-mixing conductance of the junction resulting from the different phases picked up by the electrons upon reflecting the interface. 
 Here $G_q = e^2/h$ is the conductance quantum and we sum over all the transmission channels.
 Note that as shown in Ref.\onlinecite{Eschrig2015}
 the spin-mixing term in Eq.(\ref{bc_SF}) in general is characterized by the spin quantization axis non-collinear with that of the spin-filtering term $\bm m\nparallel \bm h$. 
 For low-transparent S/F interfaces  we can safely neglect the proximity effect in the ferromagnet and use $\check g_F = {\rm sgn} \omega \hat\tau_z$ in Eq.(\ref{bc_SF}). 
 This assumption is correct provided the 
 $G_T\ll \sigma_S/\xi_S$ where $\xi_S=\sqrt{D/2\pi T_c}$ 
 is the coherence length in the S region and $T_c$ is the superconductor critical temperature. Under these conditions the thickness of the F layer in this model can be arbitrary. 
 
 The Usadel equation in the superconducting film reads:
\begin{align}
- D \bm \nabla \left( \check g  \bm \nabla \check g \right) +   
\left[ \omega \hat\tau_z  + i\check \Delta (\bm r), \check g \right]=0,
\label{usadel_S}
\end{align}
 
Due to the condition $d < \xi_S$  we can consider $\check g$ as spatially constant in the $z$-direction in the superconducting film.
Hence integrating Eq.~(\ref{usadel_S}) from $z=0$ to $z=d$ and using Eq.~(\ref{bc_SF}) we obtain the  effective Usadel Eq.(\ref{usadel_spin_filtering_p}) with $\Gamma = D(t^2 + u^2)$,  polarization $ \bm P = 2D ut \bm h /(h \Gamma )$ and 
the effective exchange field is determined by the spin-mixing angle $\bm h_{eff} = \bm m D G_\phi/2\sigma_S$.
In this case the gap function is not affected $\Delta_{eff}=\Delta$.  One can see also that in general $\bm h_{eff} \nparallel \bm P$ because the spin-rotation and spin-polarization axes of the interface are not necessarily  collinear \cite{Eschrig2015}. 
In general, if the S/F interface transparency is not low, the proximity effect in the ferromagnet can be essential and in this case the effective parameters $\Gamma$, $h_{eff}$ and $P$ can become dependent on the quasiparticle energy. However, even in this case it is the $P$-term that violates the symmetry $\bm j (\bm M) = \bm j (-\bm M)$ and accounts for magnetoelectric effects. 
The same generalized Usadel Eq.(\ref{usadel_spin_filtering_p}) for system in Fig.~\ref{sketch}(a) can be obtained using less general model based on the tunneling Hamiltonian approach. The corresponding derivation is shown in Appendix. 

Further we consider the second model system, which is sketched in Fig.~\ref{sketch}(b). The system includes a singlet superconductor 
at $d>z>d_F$ and  a weak ferromagnet at $d_F>z>0$ with the exchange field $h \ll \varepsilon_F$, where $\varepsilon_F$ is the Fermi energy. Both layers are thinner than the corresponding coherence lengths $d_S<\xi_S$ and $d_F< \sqrt{D/h}$. Assuming that the S/F interface is  fully-transparent we follow the approach suggested in Ref.\onlinecite{Bergeret2001} and treat the whole S/F bilayer as an effective ferromagnetic superconductor with the order parameter $\Delta_{eff} = (d_S/d)\Delta $ and the exchange field $\bm h_{eff} = (d_F/d)\bm h$. 
 In this case the bilayer is described by the standard Usadel equation:
  
\begin{align} \label{usadel_SF}
  \bm \nabla \left(D \check g \bm \nabla \check g \right)
= 
\left[ \omega \hat\tau_z  -i \bm h_{eff}(\bm r)
{\bm{\hat \sigma} \hat\tau_z}  + i\check \Delta_{eff} (\bm r), \check g \right],
\end{align}
  In principle, one can assume that the diffusion coefficient $D$ here is different in F and S layers $D(z<d_F)=D_F$ and $D(z>d_F)=D_S$.   
Eq.~(\ref{usadel_SF})  should be supplemented by the boundary conditions at $z=0$ and $z=d$. 
 As usual at the impenetrable F/vacuum interface $z=0$ we have
 $\check g \partial_z \check g|_{z=0} = 0 $.
The interface at $z=d$ is formed by the spin-filtering  barrier between the superconductor and the normal metal characterized by  different tunnel conductances for spin up and spin down electrons. 
 %
%
It is described by the boundary conditions Eq.~(\ref{bc_SF}) with $G_\phi = 0$. Such model for the spin-filtering interface has been widely used in works on SFS Josephson junctions
\cite{BergeretVerso2012, Bergeret2012b, Rabinovich2018, Silaev2017} 
and transport phenomena in  superconductors with spin-splitting field \cite{RevModPhys.90.041001}
 \begin{eqnarray}
 \label{bc_SF_Tunnel}
 \check g \partial_z \check g \bigl |_{z=d}
 =
 - [ \hat \Gamma \check g_N \hat \Gamma^\dagger, \check g] ,
 \end{eqnarray}
where the matrix $\hat \Gamma$ is the same as in Eq.(\ref{bc_SF}).
 %
Assuming that tunneling interface at $z=d$ has low transparency we neglect the proximity effect in the  normal metal layer at $z>d$. Thus the thickness of this layer can be arbitrary and the Green's function there is taken in the form $\check g_N = {\rm sgn} \omega \hat\tau_z$. Integrating Eq.~(\ref{usadel_SF}) from $z=0$ to $z=d$ with the boundary conditions (\ref{bc_SF_Tunnel}) we again arrive at Eq.(\ref{usadel_spin_filtering_p}). 
In this case the diffusion coefficient is 
also determined by the effective thickness-averaged value $D = (D_S d_S + D_F d_F)/d$.  

Having obtained generalized Usadel Eq.(\ref{usadel_spin_filtering_p}) from three different models we conclude that it is quite general result which allows describing 
the interaction of superconducting order with non-uniform magnetic textures and spin-polarized transport. 

\section{anomalous ground state phase shifts in S/F bilayers containing spin textures}
\label{phase_shifts}

Here we consider the inverse magnetoelectric effect in a S/F bilayer containing a magnetic texture. While the general consideration is valid for an arbitrary texture depending on the only spatial coordinate $x$, we focus on two particular examples of the magnetic helix and the head-to-head domain wall. 

The magnetization texture is described by
\begin{eqnarray}
\bm h = h (\cos \theta, \sin \theta \cos \delta, \sin \theta \sin \delta),
\label{h_texture_hhwall}
\end{eqnarray}
where in general the both angles depend on $x$-coordinate. 

Let's make the spin gauge transform in Eq.~(\ref{usadel_spin_filtering_p}) in order to work in the reference frame where the quantization axis is aligned with the local magnetization direction: $\check g=U \check g_l U^\dagger$ with $U^\dagger \bm h_{eff}(\bm r)\hat{\bm \sigma} U = h_{eff} \hat\sigma_z$. Then we obtain from Eq.~(\ref{usadel_spin_filtering_p}):
 \begin{eqnarray}
 - D \hat \partial_{\bm r}\bigl( \check g_l 
 \hat \partial_{\bm r} \check g_l \bigr) + 
 \Bigl[ (\omega +  {\rm sgn} \omega \Gamma) \hat\tau_z  -i  h_{eff} \hat\sigma_z \hat\tau_z  +  \nonumber \\
 {\rm sgn} \omega \Gamma \tilde {\bm P} \hat{\bm \sigma} +i\check \Delta (\bm r), \check g_l \Bigr]=0,~~~~~~
 \label{usadel_local}
 \end{eqnarray}
 where $\hat \partial_{\bm r}=\partial_{\bm r} + i \bigl[ \bm M_k^S \hat\sigma_k m_S, ... \bigr]$ is the gauge-covariant derivative with $M_{kj}^{S}={\rm Tr}[\hat\sigma_k U^\dagger \partial_j U]/2im_{S}$ and $\tilde {\bm P} \hat{\bm \sigma} = U^\dagger \bm P \hat{\bm \sigma} U$ is the interface polarization term in the local spin basis. In general $\tilde {\bm P}$ depends on the $x$-coordinate even if $\bm P$ is spatially independent.

The spin rotation is given by
\begin{eqnarray}
\hat U = e^{-i\hat\sigma_x \delta/2} e^{-i\hat\sigma_z \theta/2} e^{-i\hat\sigma_y \pi/4}.
\label{spin_rotation}
\end{eqnarray}
In this considered case when the magnetization texture only depends on the $x$-coordinate the gauge field can be written as follows:
\begin{eqnarray}
M_{kx}^S \hat\sigma_k m_S= \bm a \hat{\bm \sigma} ,
\label{SO_hh}
\end{eqnarray}
where $\bm a =\frac{1}{2} (\partial_x \theta,(\partial_x \delta) \sin \theta, - (\partial_x \delta) \cos \theta )$ is a vector in spin space. The other components of the spin gauge field $M_{ky}^S$ and $M_{kz}^S$ are zero.

First we assume that $\bm P$ is aligned with $\bm h$ and solve the effective Usadel equation (\ref{usadel_local}) in the S film. For simplicity we consider the linearized version of this equation valid near the critical temperature. The linearized Usadel equation for the anomalous Green's function $\hat f_l = f_0 \hat\sigma_0 + \bm f \hat{\bm \sigma}$ takes the form [we consider $\omega>0$, for $\omega<0$ the solutions can be obtained as $f_0(\omega)=f_0(-\omega)$, $\bm f(\omega)=-\bm f(-\omega)$]:

 \begin{align} \label{fv_1}
 & -\frac{D}{2}\partial_x^2 \bm f + (\omega+\Gamma + 2Da^2)
 \bm f + 
   2D [\bm a \times \partial_x \bm f] +
 \\ \nonumber 
 &  D [\partial_x \bm a \times \bm f] +   
 i \Gamma P [\bm e_z \times \bm f] 
 - 2D (\bm a \bm f) \bm a - i h_{eff}  f_0 \bm e_z = 0,
    \\
 & -\frac{D}{2}\partial_x^2 f_0 + (\omega+\Gamma )f_0 -  i h_{eff}  f_z - i \Delta = 0.
  \label{f0_1}
  \end{align}

 The general expression for the current reads
 \begin{align}
  j_x = 
 \frac{i\pi T}{2 e\rho_N } \sum \limits_{\omega>0} \Biggl\{ {\rm Tr}_2 \bigl[ \hat f_l \partial_x \hat {\tilde f}_l - \hat {\tilde f}_l \partial_x \hat f_l \bigr] + 8 (\bm f \times \tilde {\bm f})\bm a \Biggr\},
  \label{current_DW}
  \end{align}
where $\rho_N = 1/(2 e^2 N_F D)$ is the resistivity of the superconducting film in the normal state. 

The electric current can be represented as the sum of the ordinary $j_o$ and anomalous $j_a$ parts. The ordinary and anomalous contributions are given by the 
 first and the second terms in the curly brackets in Eq.(\ref{current_DW}). The anomalous contribution is defined as 
 the current in the absence of the phase difference $j_a = j(\partial_x \varphi = 0)$. 
Now our goal is to find the anomalous current, that is to solve Eqs.~(\ref{fv_1})-(\ref{f0_1}) at $\partial_x \varphi = 0$. We solve Eqs.~(\ref{fv_1})-(\ref{f0_1}) in the approximation of spatially slow magnetic texture with the characteristic length scale $d_W \gg \xi_S$. The solution up to the leading order in the parameters $\xi_S/d_W$ and $\Gamma P/(\omega + \Gamma)$ takes the form:
\begin{eqnarray}
f_0=\frac{i \Delta (\omega + \Gamma)}{(\omega + \Gamma)^2 + h_{eff}^2},
\label{f0_sol}
\end{eqnarray}
\begin{eqnarray}
f_z=\frac{-\Delta h_{eff} }{(\omega + \Gamma)^2 + h_{eff}^2},
\label{fz_sol}
\end{eqnarray}
\begin{eqnarray}
f_{x,y}=-\frac{D f_z}{(\omega + \Gamma)}\Bigl[ \pm \partial_x a_{y,x} - 2 a_z a_{x,y} - \nonumber
\\
\frac{i \Gamma P}{(\omega + \Gamma)} \bigl( \partial_x a_{x,y} \pm 2 a_z a_{y,x} \bigr)\Bigr].
\label{fxy_sol}
\end{eqnarray}
The function components
$\tilde f_i$ can be obtained from the corresponding expressions for $f_i$ with the substitution $\Delta \to -\Delta$, $P \to -P$ and $\varphi \to -\varphi$. It is seen from Eq.~(\ref{fxy_sol}) that $f_{x,y}$ are of the second order with respect to $\xi_S/d_W \equiv \sqrt{D/2 \pi T}/ d_W$. The components $f_{0,z}$ also contain second order in $\xi_S/d_W$ contributions, but they do not contribute to the anomalous current $j_a$.

Substituting Eqs.~(\ref{f0_sol})-(\ref{fxy_sol}) into the Eq.~(\ref{current_DW}) we obtain the following expressions for the ordinary and anomalous currents:
 
\begin{align} \label{ja}
 & j_a (x) = 
 \\ 
\nonumber
& -  \frac{8 \pi T D \Gamma P}{e\rho_N}
 \sum \limits_{\omega>0} \frac{f_z^2}{(\omega + \Gamma)^2}\Bigl[ (\partial_x \bm a \times \bm a)_z + 2 a_z (a_x^2 + a_y^2) \Bigr] 
 \\ \label{Eq:jo}
 & j_o = - \frac{2\pi T}{e\rho_N} \sum \limits_{\omega>0} [f_0^2+f_z^2] \partial_x \varphi_0
\end{align}
The ground state of the system is determined by the condition of zero total electric current $j_o + j_a =0$.  
 Using Eqs.(\ref{ja}), (\ref{Eq:jo}) we obtain the ground state that supports the gradient of superconducting phase
\begin{eqnarray}
\partial_x \varphi_0 = \frac{j_a e\rho_N}{2\pi T \sum \limits_{\omega>0} [f_0^2+f_z^2]}.
  \label{phase_der}
  \end{eqnarray}

{\it Magnetic helix.}
Now we consider the special case of magnetic texture in the form of a helix. In this case $d\theta/dx = 0$ and $d\delta /dx = 2\pi\kappa/L$, where $L$ is the spatial period of the helix and $\kappa = \pm 1$ determines its chirality.
\begin{align}
\partial_x \varphi_0 = 
8\pi^3\cos \theta \sin^2 \theta
\frac{\kappa D\Gamma P }{L^3}\left (\frac{ \sum \limits_{\omega>0} f_z^2/(\omega+\Gamma)^2}
{\sum \limits_{\omega>0} [f_0^2 + f_z^2]} \right)
\label{phase_helix}
\end{align}
One can see that in this case the ground state of the superconductor corresponds to the helical state - the superconducting state with zero supercurrent and a constant phase gradient $\bm \nabla \varphi = \partial_x \varphi_0 \bm e_x$. Earlier the helical state has already been predicted for superconducting systems with intrinsic spin-orbit coupling and under a uniform Zeeman field \cite{Edelstein1989,Barzykin2002,Samokhin2004,Kaur2005,Dimitrova2007,Houzet2015}. Here we report that this state can be also realized in S/F spin-textured bilayers without an intrinsic spin-orbit coupling. It is important that the helical state can only appear for {\it conical} ferromagnets where the magnetization is noncoplanar. In case of a spiral ferromagnet with $\theta = \pi/2$ the anomalous phase gradient is zero as can be seen from Eq.~(\ref{phase_helix}). Indeed, the exact solution for the superconducting state in the presence of the spiral exchange field has been investigated\cite{Bulaevskii1980} and only homogeneous superconductivity was predicted. It is also worth to note here that, while looking quite similar, this state is in sharp contrast to the famous Fulde-Ferrel-Larkin-Ovchinnikov (FFLO) state \cite{Larkin1965,Fulde1965}, where the direction of the phase gradient is not fixed by the exchange field. Here the direction of the phase gradient is strictly fixed by the magnetization texture.  

Introducing the orthogonal vectors ( $\hat {\bm h} \equiv \bm h/h$):
\begin{align} \label{Eq:n_delta}
 & \bm n_\delta =-\partial_\delta \hat {\bm h}/\sin \theta
 \\
 \label{Eq:n_theta}
& \bm n_\theta =-(\partial_x \bm n_\delta \partial_\theta \hat {\bm h})
\partial_\theta \hat {\bm h},
\end{align}
 which are also orthogonal to $\bm h$, we can also rewrite Eq.~(\ref{phase_helix}) in the form:
\begin{eqnarray}
\partial_x \varphi_0 = 
 4\pi^2 \chi_{int} P\sin^2 \theta\frac{D\Gamma}{L^2}
 \left(\frac{\sum \limits_{\omega>0} f_z^2/(\omega+\Gamma)^2}{\sum \limits_{\omega>0} [f_0^2 + f_z^2]} \right),
\label{phase_helix_inv}
\end{eqnarray}
where $\chi_{int} = \hat {\bm h} (\bm n_\theta \times \bm n_\delta) = (2\pi \kappa/L) \cos \theta$ is the invariant, describing the internal chirality of the helix. $P\chi_{int}$ is the same chiral invariant which was introduced in Ref.~\onlinecite{Rabinovich2018} to describe the anomalous Josephson effect in S/F/S junctions with a helix magnetic interlayer. It illustrates the universality of the chiral nature of the inverse magnetoelectric effect in superconducting hybrids with textured ferromagnets.

\begin{figure}[!tbh]
 \begin{minipage}[b]{\linewidth}
   \centerline{\includegraphics[clip=true,width=3.2in]{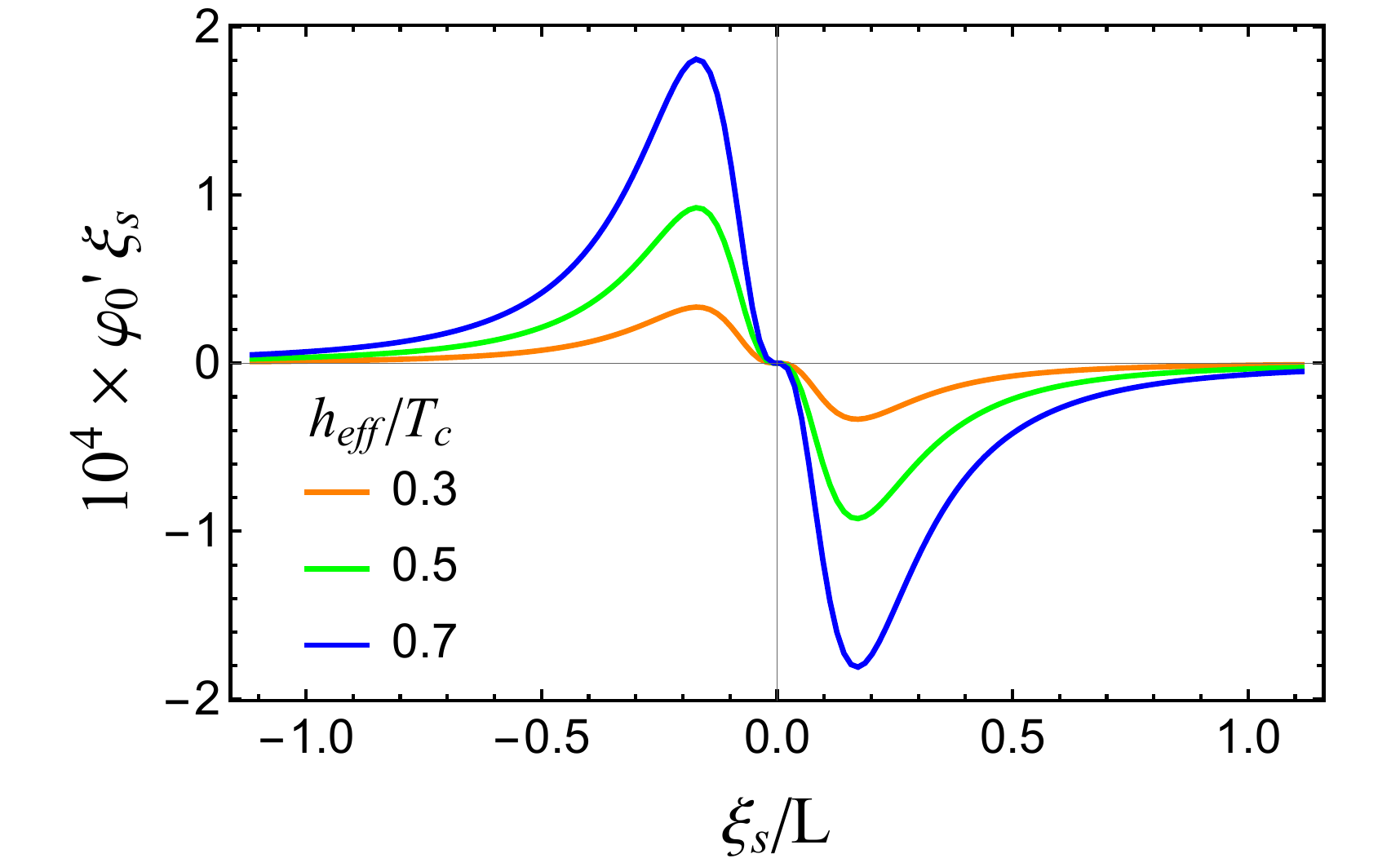}}
   \end{minipage}
   \begin{minipage}[b]{\linewidth}
   \centerline{\includegraphics[clip=true,width=3.2in]{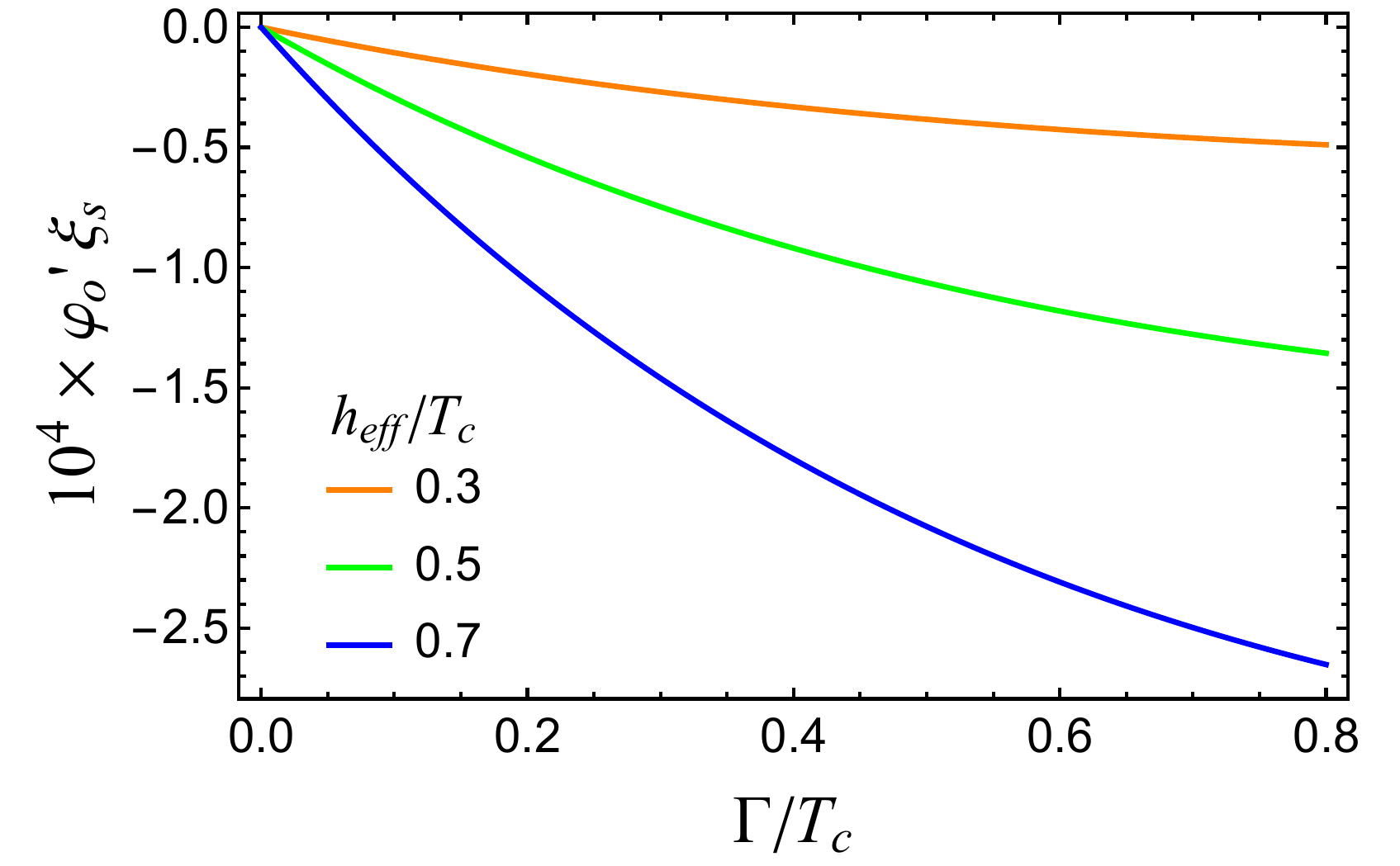}}
   \end{minipage}
   \caption{(a) $\partial_x \varphi_0$ as a function of the inverse helix period $L^{-1}$ for different $h_{eff}$. $\Gamma = 0.4T_c$.  (b) $\partial_x \varphi_0$ as a function of $\Gamma$ for different $h_{eff}$. $2\pi \xi_S/L = 1$. For the both panels $P=0.5$, $\theta = \pi/3$.}
 \label{helix}
 \end{figure}

Eqs.~(\ref{phase_helix}) and (\ref{phase_helix_inv}) describe the helical state in the limit of slow helices $\xi_S/L \ll 1$. The dependence of $\partial_x \varphi_0$ on the inverse helix period in more general case of arbitrary $L$ (still larger than the atomic scales) is presented in Fig.~\ref{helix}. It is seen that that maximal values of the phase gradient are reached for helices with the period $L \sim 2\pi \xi_S$. The nonmonotonic dependence of $\partial_x \varphi_0 (L^{-1})$ can be understood as follows. At $L^{-1} \to 0$ the value of the effective spin-orbit coupling $a$ also tends to zero. In this case the anomalous current at zero phase gradient vanishes, as it is seen from Eq.~(\ref{current_DW}). Therefore the reason for the phase gradient disappears. In the opposite limit $\xi_S/L \gg 1$ the helix magnetization oscillates too rapidly in the $(y,z)$-plane and averages to zero value at the spatial scales $\sim \xi_S$. The superconductor cannot respond at scales much smaller than $\xi_S$ and, therefore, anomalous Green's functions behave as in the homogeneous exchange field along the $x$-axis. In this case the exchange field is coplanar and magnetoelectric effects should vanish. 

Fig.~\ref{helix}(b) also demonstrates that the stronger the spin-dependent part of the depairing $P \Gamma$ the larger the phase gradient of the helical state. In fact, the dependence $\partial_x \varphi_0 (\Gamma)$ is nonmonotonous and $|\partial_x \varphi_0|$ is strongly reduced at $\Gamma >T_c$. But this part of the curve $\partial_x \varphi_0 (\Gamma)$ is not plotted because the superconductivity by itself is already suppressed at such strong depairing factors.  

{\it Domain wall.}
Further we consider a domain wall case.
Previous consideration within the quasiclassical theory did not find spontaneous phase differences\cite{Buzdin2011}. 
For definiteness we focus on the head-to-head domain wall with $\theta (x) \to 0,\pi$ at $x \to \mp \infty$. 

If the wall is coplanar, what is equivalent to the condition $\partial_x \delta = 0$, then $j_a = 0$, as it can be easily seen from Eq.~(\ref{ja}) and the definition of vector $\bm a$. In this case the magnetoelectric effect is absent and the ground state of the superconductor is a homogeneous state with a spatially constant phase.

If the DW is characterized by non-coplanar magnetization distribution, that  is $\partial_x \delta \neq 0 $, then $j_a \neq 0$ and between any points $x_1$ and $x_2$ of the superconductor in the ground there is a phase difference, which is to be calculated as $\int \limits_{x1}^{x_2} \partial_x \varphi_0 (x)dx$, where $\partial_x \varphi_0$ is defined by Eq.~(\ref{phase_der}). It is obvious that this phase difference is zero far from the DW, but is finite if the DW is located inside the region between $x_1$ and $x_2$. The noncoplanarity of the wall can be caused by different reasons. For example, it can be induced by the supercurrent moving the DW, or it can be just due to the contact between the ferromagnet and the superconductor, because in this case it is energetically more favorable to disturb the initial wall texture in order to reduce stray fields penetrating the superconductor. In any of the described cases the resulting magnetization structure and the ground state phase difference can be calculated, but this problem is beyond the scope of the present work. Instead, here we consider the case of the externally induced noncoplanarity in the system and demonstrate that in this case the resulting ground state phase difference is also governed by a  chiral invariant.  

The simplest model system where the external (not caused by the internal chirality of the ferromagnet texture)  chiral invariant takes place is sketched in Fig.~\ref{sketch}(b). It is assumed that the wall is coplanar, but the polarization $\bm P$ of the spin-filtering interface is not fully aligned with the ferromagnet magnetization. 

In order to obtain nonzero anomalous current $j_a$ for the considered case of external chirality  it is enough to find the anomalous Green's functions up to the zero order in $\xi_S/d_W$. Then the linearized Usadel equations for the anomalous Green's function take the form:
 \begin{align}\label{fv_2}
 & \frac{D}{2}\partial_x^2 \bm f - (\omega+\Gamma )\bm f + 
 i h_{eff}  f_0 \bm e_z - i \Gamma  [\tilde {\bm P} \times \bm f] = 0,
 \\  \label{f0_2}
 & \frac{D}{2}\partial_x^2 f_0 - (\omega+\Gamma )f_0 +  i h_{eff} f_z + i \Delta = 0.
   \end{align}

The solution of these equations up to the first order with respect to $|\tilde {\bm P}(x)|\Gamma/(\omega+\Gamma)$ can be written in the compact form:
\begin{eqnarray}
\bm f_\perp = \frac{-i\Gamma f_z}{(\omega+\Gamma)}[\tilde {\bm P} \times \bm e_z],
\label{fperp_sol}
\end{eqnarray}
where $\bm f_\perp = (f_x, f_y, 0)$ is the component of the triplet anomalous Green's function $\bm f$ in plane perpendicular to the local quantization axis. The components $f_z$ and $f_0$ are expressed by Eqs.~(\ref{fz_sol}) and (\ref{f0_sol}), respectively.

For the coplanar domain wall $\bm a = (1/2)(d\theta/dx) \bm e_x$ and the anomalous current can be expressed as
\begin{eqnarray}
j_a (x) =
\frac{4 \pi T \Gamma}{e\rho_N} \sum \limits_{\omega>0} \frac{f_z^2}{\omega + \Gamma}\bigl(\bm P[\partial_x \hat {\bm h} \times \hat {\bm h}] \bigr).
\label{ja_dw}
\end{eqnarray}

The corresponding ground state phase gradient can be found according to Eq.~(\ref{phase_der}). The total phase difference acquired in the superconductor due to the DW presence can be found as $\Delta\varphi_0 = \int \limits_{-\infty}^\infty \partial_x \varphi_0 dx$ and takes the form:
\begin{eqnarray}
\Delta \varphi_0 = 2\pi \Gamma |P_\perp| \chi_{ex} \frac{
 \sum \limits_{\omega>0} f_z^2/(\omega + \Gamma)}{\sum \limits_{\omega>0} [f_0^2+f_z^2]},
\label{phase_DW}
\end{eqnarray}
where $\chi_{ex}={\rm sgn}\left[\bm P(\partial_x \hat {\bm h}\times \hat {\bm h}) \right]$, $P_\perp$ is the component of the polarization vector $\bm P$ perpendicular to the DW plane. We see that the ground state phase difference acquired by the superconductor due to the presence of the DW in the ferromagnet is controlled by the external chirality invariant $\chi_{ex}$, which is only nonzero if the polarization $\bm P$ of the spin-filtering interface has the component perpendicular to the wall plane. 

\begin{figure}[!tbh]
 \begin{minipage}[b]{\linewidth}
   \centerline{\includegraphics[clip=true,width=3.2in]{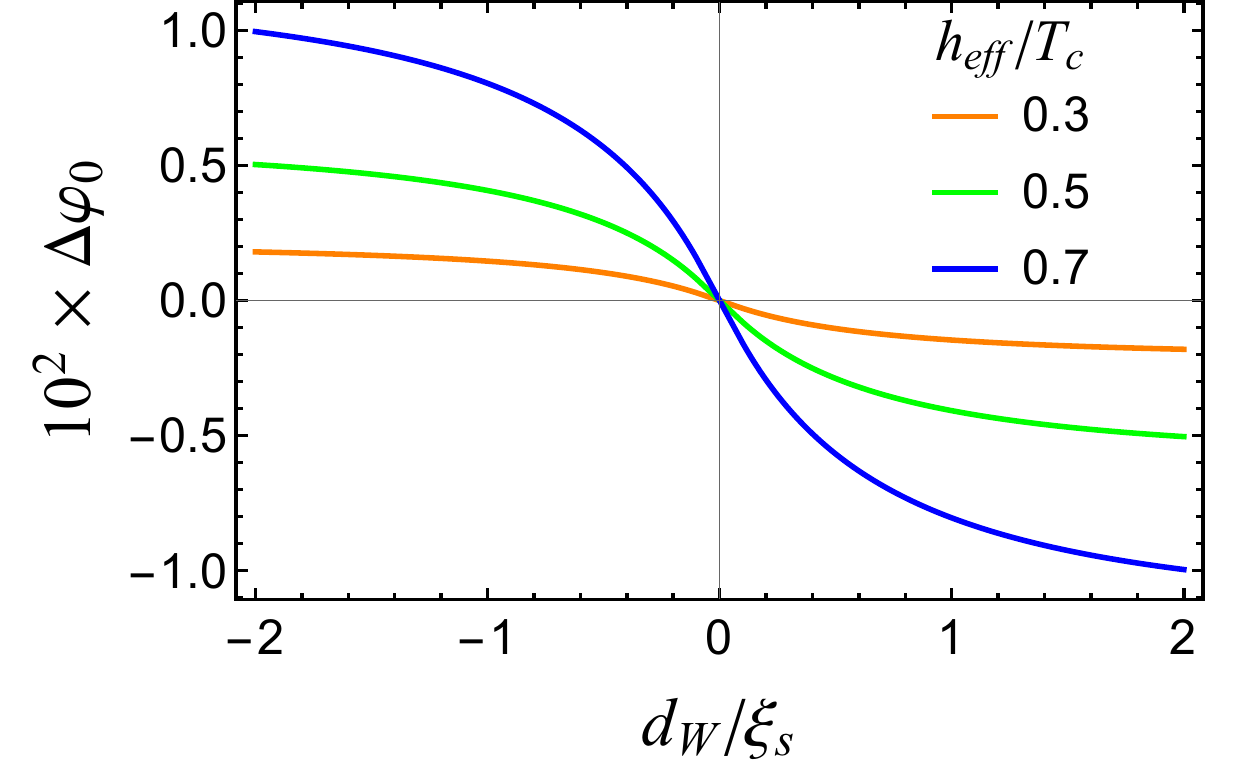}}
   \end{minipage}
   \begin{minipage}[b]{\linewidth}
   \centerline{\includegraphics[clip=true,width=3.2in]{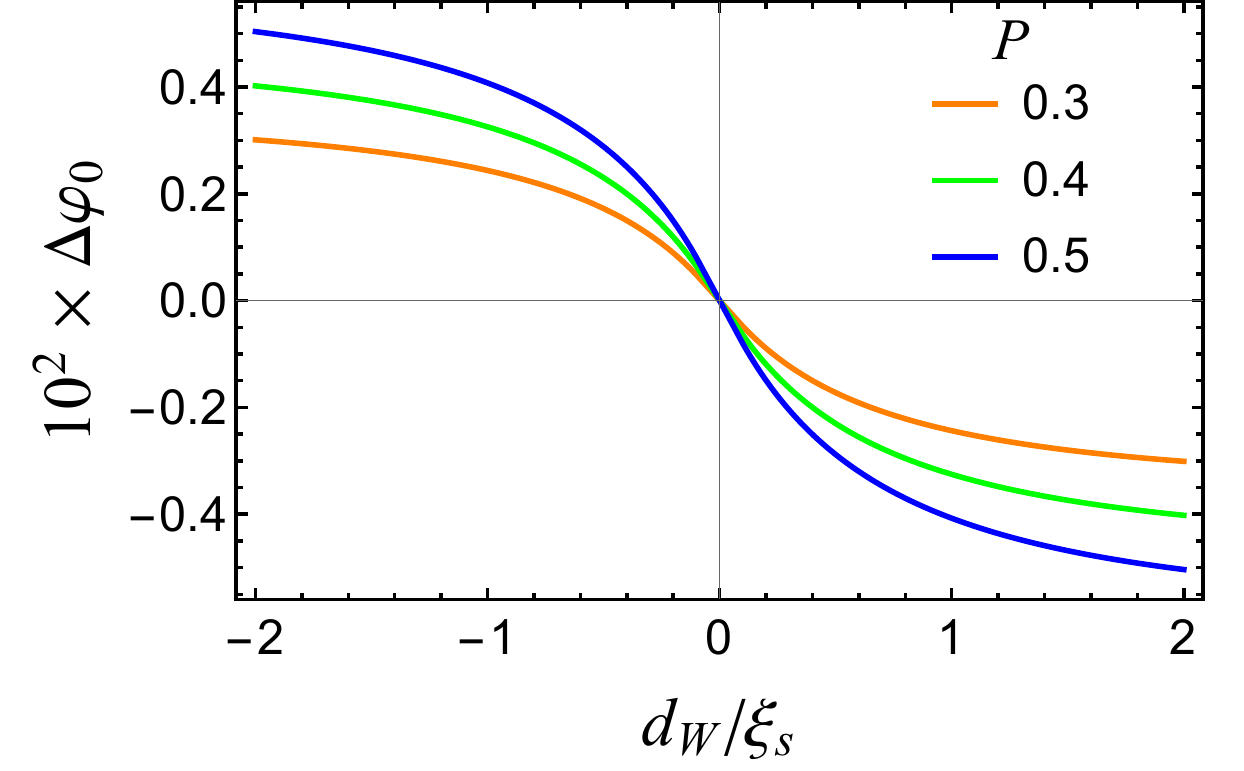}}
   \end{minipage}
   \caption{Phase difference acquired at the DW in case of external chirality (see text) as a function of $d_W^{-1}$. (a) Different curves correspond to different $h_{eff}$, $P=0.5$. (b) Different curves correspond to different $P$, $h_{eff} = 0.5 T_c$. For the both panels $\Gamma = 0.4 T_c$.}
 \label{DW}
 \end{figure}

Eq.~(\ref{phase_DW}) gives the ground state phase difference at the domain wall in the limit of wide DW $d_W \gg \xi_S$. Beyond this limit our analytical treatment is inapplicable, but the result of the numerical calculation is presented in Fig.~\ref{DW}. It is seen that the maximal phase difference is acquired at wide walls with $d_W \gg \xi_S$, when $\Delta \varphi_0$ tends to the analytical answer expressed by Eq.~(\ref{phase_DW}).  

In principle, there can be also a contribution to the anomalous current from the vector potential of the stray fields generated by the DW. However, this contribution depends on the $y$-coordinate and should be zero after averaging over it. Therefore, the averaged over the $y$-coordinate phase difference is still given by considered magnetoelectric effect even if the stray fields are taken into account.
 
In general the discussed above  phase difference arising in a superconductor due to the magnetoelectric effect provides the connection between the the magnetic texture and the condensate phase. In particular, it opens a way to detect electrically time-dependent textures (for example, moving domain wall and other magnetic defects) via the relation $V=(\hbar/2e)\partial_t \Delta \varphi$. It is interesting that such an electrical detection resolves not only the defect movement, but also its chirality. The detailed investigation of this problem is beyond the framework of this work and we postpone it for a future study.

\section{Direct magnetoelectric effect in S/F bilayers and the related torque}
\label{Sec:torque}

Here we consider the direct magnetoelectric effect in S/F bilayers, that is the generation of an equilibrium spin polarization in response to a supercurrent. The induced polarization is found to have a component perpendicular to the ferromagnet magnetization and, therefore, gives rise to a torque acting on the ferromagnet texture $\bm M(\bm r)$. Further we study the torque in details.

Our first goal is to find the supercurrent-induced spin polarization in the superconductor $\bm m = -2\mu_B \bm s $, where $\bm s$ is an electron spin. In terms of the linearized quasiclassical Green's function in the fixed spin basis it can be calculated as follows:
\begin{eqnarray}
\bm m = \frac{-i\pi \mu_B N_F}{2} T \sum \limits_{\omega>0}  {\rm Tr}_2 \Bigl[ \hat{\bm \sigma} (\hat f \hat {\tilde f} + \hat {\tilde f} \hat f) \Bigr].
\label{spin_direct}
\end{eqnarray}
If we define the spin vectors $\bm \alpha^i$ for $i=x,y,z$ as $\bm \alpha^i \hat{\bm \sigma} = U^\dagger \hat\sigma_i U$, then in terms of the quasiclassical Green's function in the local spin basis Eq.~(\ref{spin_direct}) can be written as follows:
\begin{eqnarray}
\bm m_i = -2i\pi \mu_B N_F T \sum \limits_{\omega>0} \bm \alpha^i (f_0 \tilde {\bm f} +\tilde f_0 \bm f) .
\label{spin_direct1}
\end{eqnarray}
In order to calculate the induced electron magnetization according to Eq.~(\ref{spin_direct1}) we have to find the anomalous Green's functions in the local spin basis up to the first order in the applied supercurrent, or in other words, up to first order in the superconducting phase gradient $\partial_x \varphi$. We start from the linearized Usadel equations, valid for the case of an arbitrary directed polarization of the spin-filtering interface: 

\begin{align} \nonumber
 & \frac{D}{2}\partial_x^2 \bm f -
 (\omega+\Gamma + 2Da^2)\bm f -
  D \Bigl([\partial_x \bm a \times \bm f] + 2 [\bm a \times \partial_x \bm f] \Bigr) -
  \\  \label{fv_direct}
 &  i \Gamma [\tilde{ \bm P} 
 \times \bm f] + 2D (\bm a \bm f) \bm a = 
 -i h_{eff} f_0 \bm e_z,
   \\  \label{f0_direct}
& \frac{D}{2}\partial_x^2 f_0  + i h_{eff}  f_z + i \Delta(x) = (\omega+\Gamma )f_0 .
   \end{align}

 Assuming that $\Delta (x) = \Delta e^{i\varphi(x)}$ and performing the transformation $\hat f_l = \hat f_h e^{i\varphi(x)}$, up to the first order in $\partial_x \varphi$ and up to the first order in $\xi_S/d_W$ Eqs.~(\ref{fv_direct})-(\ref{f0_direct}) take the form
 
 \begin{align}    \nonumber
 & \frac{D}{2}\partial_x^2 \bm f_h - (\omega+\Gamma)\bm f_h 
 - 2iD\partial_x \varphi [\bm a \times \bm f_h] -  
  i \Gamma [\tilde{ \bm P} \times \bm f_h] = 
 \\ \label{fv_direct1}
 &  -i h_{eff}  f_{h0} \bm e_z,
 \\ \label{f0_direct1}
 & \frac{D}{2}\partial_x^2 f_{h0} - (\omega+\Gamma )f_{h0}  + i \Delta = - i h_{eff}  f_{hz}
    \end{align}

 The solution of these equations up to the first order in $P\Gamma/(\omega+\Gamma)$ takes the form

 \begin{align} \label{fz_sol_direct}
 & f_{hz}=f_z\Bigl( 1+\frac{4 D \partial_x \varphi \Gamma} 
 {(\omega+\Gamma)^2 + h_{eff}^2}\bm a_\perp \tilde 
 {\bm P}_\perp \Bigr),
 \\ \label{f0_sol_direct}
 & f_{h0}=f_0+\frac{i h_{eff} }{(\omega+\Gamma)}
 \frac{4 D \partial_x \varphi \Gamma f_z}{(\omega+\Gamma)^2 
 + h_{eff}^2} \bm a_\perp \tilde {\bm P}_\perp ,
 \\ \label{fxy_sol_direct}
 & \bm f_{h\perp}=\bm f_\perp - \frac{2 i \partial_x \varphi  
 D[\bm a \times \bm e_z]f_z}{\omega+\Gamma}-
 \\ \nonumber
 & \frac{2  \partial_x \varphi D \Gamma f_z} 
 {(\omega+\Gamma)^2}\bigl[ a_z \tilde {\bm P}_\perp +
 \tilde P_z \bm a_\perp \bigr],
 \end{align}
 %
where $f_z$, $f_0$ and $\bm f_\perp$ are defined by Eqs.~(\ref{fz_sol}), (\ref{f0_sol}) and (\ref{fperp_sol}), respectively. $\tilde {\bm P}_\perp$ is the component of the vector $\tilde {\bm P}$, which is perpendicular to the local direction of the ferromagnet magnetization. $\bm a_\perp$ is defined in the same way. The anomalous Green's function ${\hat {\tilde f}}_h$ can be obtained from Eqs.~(\ref{fz_sol_direct})-(\ref{fxy_sol_direct}) with the substitution $\Delta \to -\Delta$, $P \to -P$ and $\partial_x \varphi \to -\partial_x \varphi$.

Substituting the anomalous Green's functions into Eq.~(\ref{spin_direct1}) we obtain the following result for the supercurrent-induced electron magnetization:
\begin{eqnarray}
 \bm m = 4\pi \mu_B N_F T D \partial_x \varphi  \sum  \limits_{\omega>0} \frac{f_z}{(\omega+\Gamma)  [(\omega+\Gamma)^2 + h_{eff}^2]} \times 
 \nonumber \\
 \Gamma \Delta \Bigl\{ (\partial_x \hat {\bm h} 
 \times \hat  {\bm h}  )(\bm P \hat {\bm h}) - 
 \delta^\prime \cos \theta \tilde {\bm P}_\perp \Bigr\}.~~~~~~~~~~ 
\label{spin_final}
\end{eqnarray}
Here we have only written the magnetization component, which is perpendicular to the local direction of the ferromagnet $\hat {\bm h}$ because it is this component that gives rise to a torque $\bm N$ acting on the ferromagnet magnetization. The torque can be calculated as follows: 
\begin{eqnarray}
\bm N = 2 h \beta (\hat {\bm h} \times \bm m),
\label{torque}
\end{eqnarray}
where $\beta = h_{eff}/h <1$ is the dimensionless coefficient between the actual exchange field of the ferromagnet $h$ and the effective exchange field $h_{eff}$, which is induced in the superconductor due to the magnetic proximity effect. Substituting $\bm m$ from Eq.~(\ref{spin_final}) the torque can be written as

\begin{align} \label{torque1}
& \bm N = b_j \partial_x \hat {\bm h} + c_j \chi_{int} (\tilde {\bm P}_\perp \times \hat {\bm h}),
\\ \label{cj}
& c_j = \pi \mu_B N_F T D \partial_x \varphi \sum \limits_{\omega>0} \frac{8 \beta h\Gamma \Delta f_z}{(\omega+\Gamma)[(\omega+\Gamma)^2 + h_{eff}^2]},~~~~ 
\\ \label{bj}
& b_j = c_j (\bm P \hat {\bm h}), 
\end{align}

Here $\chi_{int}$ is the local internal chirality of the ferromagnet texture, defined in the same way as for the magnetic helix case:
\begin{eqnarray}
\chi_{int} = \delta^\prime \cos \theta = \hat {\bm h} (\bm n_\theta \times \bm n_\delta),
\label{bj}
\end{eqnarray}
with $\bm n_\delta$ and $\bm n_\theta$ defined by Eqs.~(\ref{Eq:n_delta})-(\ref{Eq:n_theta}).
The expression for coefficient $c_j$ (\ref{cj}) can be rewritten in terms of the supercurrent flowing via the superconductor:

\begin{align} \label{cj_current}
& c_j =-2 \beta \Gamma \Delta  \mu_B h j  \times
\\ \nonumber
& \sum \limits_{\omega>0} 
\frac{f_z}
{(\omega+\Gamma)[(\omega+\Gamma)^2 + h_{eff}^2]}
\bigg/ \sum \limits_{\omega>0}(f_{h0}^2+f_{hz}^2),~~~~ 
\end{align}
%
where up to the leading order in $\xi_S/d_W$ we have neglected the anomalous current $j_a$.

The first term in Eq.~(\ref{torque1}) represents the adiabatic spin transfer torque (STT).
Typically the adiabatic STT is determined by the transfer of the angular momentum from the current-carrying electrons to the ferromagnet magnetization. Thus coefficient $b_j$ is proportional not only to the electric current $j$, but also the degree of its spin polarization. 

Here the microscopic origin of the adiabatic STT is different. First let us consider the case when $\bm P$ is aligned with the ferromagnet magnetization. Then $\tilde {\bm P}_\perp = 0$ and the adiabatic STT is the only contribution to the torque in the system. It can be demonstrated that in this case the spin current through the system is zero to the considered accuracy. Therefore, the electric current is not spin polarized and the torque  is not connected to the derivative of the spin current, that is to the spin transfer from the current-carrying electrons to the magnetization. Its mechanism is connected to the creation of current-induced spin-resolved DOS in the superconductor in the region contacted to the textured area of the ferromagnet. Therefore, it is specific only for superconducting systems and can also be relevant as for hybrid superconducting systems with ferromagnetic metals, so as for hybrids with ferromagnetic insulators. 

If $\bm P$ is not fully aligned with the ferromagnet magnetization, the other part of the torque, expressed by the second term in Eq.~(\ref{torque1}) can appear. In general, it has components as along the direction $\partial_x \hat {\bm h}$, so as along the perpendicular direction $\hat {\bm h} \times \partial_x \hat {\bm h}$. But it cannot be included neither to the adiabatic STT, nor to the non-adiabatic STT, because it is proportional to the local internal chirality of the structure $\chi_{int}$ and, consequently, vanishes for coplanar ferromagnetic textures.

\section{Conclusions}
\label{conclusions}

We have studied the direct and inverse magnetoelectric effects in thin film S/F bilayers with spin-textured ferromagnets. The generalized Usadel equation, which allows for description of the magnetoelectric effects, is formulated. 
The inverse magnetoelectric effect leads to the formation of the phase-inhomogeneous ground state in the superconducting film  due to the exchange interaction of spin-triplet Cooper pairs with non-coplanar magnetic texture.  This effect can be used for electrical detection of DW motion. 

The direct magnetoelectric effect induces a stationary spin polarization of the superconducting  condensate in the presence of the applied supercurrent and the non-coplanar spin texture. The direction of induced Cooper pair spin is non-collinear with the local magnetization in the adjacent ferromagnetic layer. Therefore exchange interaction of the induced electron spin and the ordered magnetic moments acts as a spin torque on the ferromagnet magnetization. This torque consists of two parts. The first one is similar to usual adiabatic spin-transfer torque, and the second one is connected to the local chirality of the magnetic texture.
The chirality-sensitive term in the spin torque can be mediated only by the spin-triplet superconducting correlations since it is generically  absent in the normal state. 
The found superconducting spin torque in S/F bilayers can be used in spintronics for the low-dissipative electric current-controlled manipulation with positions of domain walls and magnetic skyrmions. 

 {\it Note added}: after this work was submitted for publication the related work considering the inhomogeneous phase state in S/F bilayers with magnetic helix was published \cite{Meng2019}
\section{Acknowledgements}

We acknowledge the financial support by the RFBR projects No. 18-52-45011 (I.V.B. and A.M.B.), No. 18-02-00318 and No. 19-02-00466 (D.S.R., I.V.B. and A.M.B.). M.A.S. was supported by the Academy of Finland Project No. 297439.

\appendix
\section{Derivation of the generalized Usadel equation in the framework of tunneling hamiltonian approach}

Here we assume that the S/F bilayer consists of a conventional superconducting film and a strong ferromagnet with the exchange field $h \sim \varepsilon_F$, as depicted in Fig.~\ref{sketch}(a). Further our goal is to derive the effective Eilenberger equation in the S film reducing the influence of the ferromagnet to the effective self-energy terms in this equation.

  The Hamiltonian of the system takes the form:
 \begin{equation}
 \hat H = \int_F d\bm r \hat H_F + \int_S d\bm r \hat H_S +
 \int_{I} d\bm r \hat H_T
 \enspace ,
 \label{full_ham}
 \end{equation}
 where the first, second  integrals are taken over the ferromagnetic, superconducting volumes, respectively. The third term describes tunneling and the integral there is taken over the interface assumed to be located at $z=0$ plane. The ferromagnet
 can be either a three-dimensional one occupying the space at $z<0$ or the two-dimensional locating at the same plane as the interface.
 The Hamiltonians of the each region are given by
 \begin{align}  \label{ham_ferr3D}
 & \hat H_F =
 \hat \psi^\dagger \Bigl[ -\frac{\nabla^2}{2m_F}-\mu - \bm h(\bm r)\hat{\bm \sigma} +
 V_{i} \Bigr]\hat \psi
 \\ \label{ham_super}
 & \hat H_S = \hat \psi^\dagger \Bigl[ -\frac{\nabla^2}{2m_S}-\mu + V_{i} \Bigr]\hat \psi +
 (\Delta\psi_{\uparrow}^\dagger \psi_{\downarrow}^\dagger+ hc)
 \\
 & \hat H_T = \hat \psi^\dagger(z=-0)  \hat t \hat \psi (z=+0) + hc
 \enspace .
 \label{ham_tunn}
 \end{align}

 Here the operators $\hat \psi(z=\pm 0)$ are taken on the superconductor/ferromagnet side respectively and  $V_{i}$ is the random impurity potential.
  We consider the simplest model, where the hopping elements are momentum-independent, but they are assumed to be spin-dependent and parametrized as $\hat t =  t_i \hat\sigma_i$.
 The thin superconducting film is assumed to be effectively two-dimensional. 
 
  We construct Green's function using the  spin-Nambu bispinors
\begin{equation}
 \hat \Psi = \left( \hat \psi_\uparrow, \hat\psi_\downarrow,
 -\hat\psi_\downarrow^\dagger, \hat\psi_\uparrow^\dagger  \right)^T
\end{equation}

Let's introduce the retarded Green's functions in the superconductor and ferromagnet:
  \begin{equation}
  \check G_{S,F}(\bm r,\bm r',t,t')= -i\theta(t-t')\hat\tau_z\langle\{\hat \Psi_{S,F} (\bm r),\hat \Psi_{S,F} ^{\dagger}(\bm r')\}\rangle,
  \label{GS_comp}
 \end{equation}
 and the tunnel Green's function
 \begin{equation}
  \check G_{T}(\bm r,\bm r',t,t')= -i\theta(t-t')\hat\tau_z\langle\{\hat \Psi_{F} (\bm r),\hat \Psi_{S} ^{\dagger}(\bm r')\}\rangle.
\end{equation}

 Then the coupled Gor'kov equations for $\check G_s$ and $\check G_T$ take the form:
 \begin{eqnarray}
 \bigl[\varepsilon \hat\tau_z + \frac{\nabla^2}{2m_S}+\mu - V_{i}^S - \check \Delta (\bm r) \bigr]\check G_S- \nonumber \\
 \check t \check G_T = \delta (\bm r - \bm r')
 \enspace ,~~~~~~
 \label{gorkov_super_new}
 \end{eqnarray}
 \begin{eqnarray}
  \bigl[\varepsilon \hat\tau_z + \frac{\nabla^2}{2m_F}+\mu - V_{i}^F + \bm h (\bm r) \hat{\bm\sigma} \hat\tau_z
  \bigr]\check G_T- \check t \check G_S = 0
 \enspace ,~~~~
 \label{gorkov_tunn_new}
 \end{eqnarray}
  where $\check \Delta(\bm r)=\Delta \hat\tau_+ - \Delta^* \hat\tau_- $, and $\check t=\hat t (1+\hat\tau_z)/2+\hat\sigma_y \hat t^* \hat\sigma_y(1-\hat\tau_z)/2 $.

 To work in the reference frame where the quantization axis is aligned with the local magnetization direction let's make the spin gauge transformation in Eqs.~(\ref{gorkov_super_new}) and (\ref{gorkov_tunn_new}): $\check G_{S,T}=U \check G_{S,T,l} U^\dagger$ with $U^\dagger \bm h(\bm r)\hat{\bm \sigma} U = h \hat\sigma_z$. Then we obtain from Eq.~(\ref{gorkov_tunn_new}):
 \begin{eqnarray}
  \bigl[\varepsilon \hat\tau_z + \frac{\nabla^2}{2m_F}+\mu - V_{i}^F + h \hat\sigma_z \hat\tau_z+ \nonumber \\
  i M_{kj}^F \hat\sigma_k \partial_j
  \bigr]\check G_{T,l} - \check t_l \check G_{S,l} = 0
 \enspace ,~~~~
 \label{gorkov_tunn_rot}
 \end{eqnarray}
 where $M_{kj}^{S,F}={\rm Tr}[\hat\sigma_k U^\dagger \partial_j U]/2im_{S,F}$ and $\check t_l = U^\dagger \check t U$ is the hopping matrix in the local basis. Further we make a natural assumption that in this {\it local spin basis}
  $\hat t_l = t_\uparrow (1+\hat\sigma_z)/2 + t_\downarrow (1-\hat\sigma_z)/2$
   is a diagonal matrix. Physically it means that the interface has only spin-filtering properties, but does not rotate spin. Additionally we assume $t_{\uparrow,\downarrow}$ to be real. In this case $\check t_l = \hat t_l (1+\hat\tau_z)/2 + \hat\sigma_y \hat t_l \hat\sigma_y (1-\hat\tau_z)/2$.

 We also can perform the impurity averaging in Eq.~(\ref{gorkov_tunn_rot}), what results in appearing of the impurity self-energy  $\check \Sigma^{S,F}(\bm r)=(1/2\pi \tau N_F)\check G_{S,F}(\bm r, \bm r, z,z)$ in the Gor'kov equations. Then denoting $H_0=\varepsilon \hat\tau_z + \frac{\nabla^2}{2m_F}+\mu - \check \Sigma^F + h \hat\sigma_z \hat\tau_z - M_{kj}^F \hat\sigma_k \partial_j $ and introducing the ferromagnet Green's function as a solution of the equation:
  \begin{eqnarray}
H_0(\bm r,z)\check G_{F,0}(\bm r,\bm r',z,z')=\delta(\bm r - \bm r')\delta(z-z')
 \enspace ,~~~~
 \label{gorkov_ferr_bare}
 \end{eqnarray}
 we obtain
 \begin{eqnarray}
  \check G_{T,l}(\bm r,\bm r')= \nonumber \\
  \int d^2 r'' \check G_{F,0}(\bm r,\bm r'',z=z'=0)\check t_l \check G_{S,l}(\bm r'', \bm r')
 \enspace .
 \label{G_tunn}
 \end{eqnarray}
 For a 2D ferromagnet the dependence on the $z$-coordinate should be omitted.

 Substituting Eq.~(\ref{G_tunn}) into the spin gauge transformed and averaged over the impurities version of Eq.~(\ref{gorkov_super_new}), we obtain
 \begin{eqnarray}
  \bigl[\varepsilon \hat\tau_z + \frac{\nabla^2}{2m_S}+\mu - \check \Sigma_S + i M_{kj}^S \hat\sigma_k \partial_j -
  \check \Delta \bigr]\check G_{S,l} -  \nonumber \\
   \int d^2 r'' \check t_l \check G_{F,0}(\bm r,\bm r'',z=z'=0)\check t_l \check G_{S,l}(\bm r'', \bm r') = \nonumber \\
   \delta (\bm r - \bm r'),~~~~
  \label{gorkov_eff}
 \end{eqnarray}
 
  Performing in Eq.~(\ref{gorkov_eff}) the Fourier transformation $(\bm r - \bm r') \to \bm p$ and subtracting from Eq.~(\ref{gorkov_eff}) the analogous equation acting on $\bm r'$, to the quasiclassical accuracy we obtain:
 \begin{eqnarray}
  \frac{i \bm p}{m}\nabla_{\bm R}\check G_{S,n}(\bm R, \bm p)+\Bigl[ \varepsilon \hat\tau_z - \check \Sigma_S - M_{kj}^S \hat\sigma_k p_j - \check \Delta (\bm R),\check G_{S,l} \Bigr] \nonumber \\
  - \Bigl[ \check t_l \check G_{F,0}(\bm R, \bm p, z=z'=0)\check t_l, \check G_{S,l} \Bigr] = 0
  \enspace , ~~~~~~~~~~
  \label{eilenberger_int}
 \end{eqnarray}
where $\bm R =(\bm r + \bm r')/2$. Let's find the explicit form of $\check G_{F,0}$ from Eq.~(\ref{gorkov_ferr_bare}). We consider only zero-order terms with respect to spin gauge field $M_{kj}$ in this equation. In other words, we neglect the terms of the order of $t^2 M_{kj}$ with respect to $M_{kj}$ in Eq.~(\ref{eilenberger_int}). In this approximation for the 3D ferromagnet we obtain:
 \begin{eqnarray}
  \check G_{F,0}(\bm p, z=z'=0)=\Bigl[ \frac{-i}{|v_{\uparrow,z}|+\frac{i}{2\tau_F |p_{\uparrow,z}|}}\frac{1+\hat\sigma_z}{2}+ \nonumber \\ \frac{-i}{|v_{\downarrow,z}|+\frac{i}{2\tau_F |p_{\downarrow,z}|}}\frac{1-\sigma_z}{2} \Bigr]\frac{1+\hat\tau_z}{2}+ \nonumber \\
  \Bigl[ \frac{i}{|v_{\downarrow,z}|-\frac{i}{2\tau_F |p_{\downarrow,z}|}}\frac{1+\sigma_z}{2}+ \nonumber \\
  \frac{i}{|v_{\uparrow,z}|-\frac{i}{2\tau_F |p_{\uparrow,z}|}}\frac{1-\hat\sigma_z}{2} \Bigr]\frac{1-\hat\tau_z}{2}
  \enspace , ~~~~
  \label{Green_bulk_ferr_3D}
 \end{eqnarray}
 where $|p_{\sigma,z}|=\sqrt{2m_F(\mu+\sigma h - \frac{p_{S}^2}{2m_F})}$ and $p_S$ is the 2D momentum in the superconductor. $\tau_F$ - is the impurity scattering time in the ferromagnet.

 For the case of the 2D ferromagnet one can obtain:
 \begin{eqnarray}
  \check G_{F,0}(\bm p)=\Bigl[ G_\uparrow^F \frac{1+\hat\sigma_z}{2}+ G_\downarrow^F \frac{1-\hat\sigma_z}{2} \Bigr]\frac{1+\hat\tau_z}{2}+ \nonumber \\
  \Bigl[ \tilde G_\downarrow^F \frac{1+\hat\sigma_z}{2}+ \tilde G_\uparrow^F \frac{1-\hat\sigma_z}{2} \Bigr]\frac{1-\hat\tau_z}{2}
  \enspace , ~~~~
  \label{Green_bulk_ferr_2D}
 \end{eqnarray}
 where $G_{\uparrow,\downarrow}^F=1/(\varepsilon-\xi_F \pm h + \frac{i}{2\tau_F})$ and $\tilde G_{\uparrow,\downarrow}^F=1/(-\varepsilon-\xi_F \pm h - \frac{i}{2\tau_F})$.

 Substituting Eq.~(\ref{Green_bulk_ferr_3D}) or (\ref{Green_bulk_ferr_2D}) into Eq.~(\ref{eilenberger_int}) and integrating it near the Fermi surface of the superconductor, we obtain the following effective Eilenberger equation in the superconducting film:
 \begin{eqnarray}
  i \bm v_S \partial_{\bm r}\check g_{S,l}(\bm r, \bm p_S)+\Bigl[ (\varepsilon+i\Gamma) \hat\tau_z  +i \Gamma_z \hat\sigma_z-\check \Sigma_S - \nonumber \\
  M_{kj}^S \hat\sigma_k p_{S,j} - \check \Delta (\bm r)+h_{eff} \hat\sigma_z \hat\tau_z,\check g_{S,l} \Bigr] = 0
  \enspace ,~~~~
  \label{eilenberger_final}
  \end{eqnarray}
  where $\bm p_S$ is the 2D Fermi momentum in the superconductor. We see that the influence of the interface with the ferromagnet on the properties of the superconducting film can be described by the effective parameters $\Gamma$, $h_{eff}$ and $\Gamma_z$ and the $SU(2)$ gauge field $M_{kj}$, which results from the magnetization texture and enters Eq.~(\ref{eilenberger_final}) exactly in the same way as the linear in momentum SO coupling does. Eilenberger equation in the form of Eq.~(\ref{eilenberger_final}) is valid as for the 3D, so as for the 2D ferromagnet contacting the thin superconducting film. Physically $\Gamma$ accounts for the leakage of the superconducting correlations into the nonsuperconducting region and is still there even if we change the ferromagnet by the normal metal. $h_{eff}$ is an effective Zeeman field. $\Gamma_z$ is nonzero only for strong ferromagnets and vanishes in the zero order with respect to $h/\varepsilon_F$, when $p_\uparrow = p_\downarrow$ and $t_\uparrow=t_\downarrow$. It drops out of the equations if the ferromagnet is homogeneous and only opposite-spin pairs are generated. Therefore, it acts only on equal-spin pairs. Physically it accounts for the fact that for an interface with strong ferromagnet the leakage of the superconducting pairs into the ferromagnet is spin-dependent: $\Gamma \pm \Gamma_z \equiv \Gamma_{\uparrow,\downarrow}$ are the depairing factors (due to the leakage into the ferromagnet) for the spin-up and spin-down pairs, respectively. It is the term  that violates the symmetry $j(\bm M) = j(-\bm M)$ and provides the possibility for nonzero spontaneous electric currents and phase inhomogeneities. The particular expression for these effective parameters for the case of 3D ferromagnet are the following:
    \begin{eqnarray}
  \Gamma = \frac{1}{2}\sum \limits_\sigma\frac{t^2_{\sigma}}{|v_{\sigma,z}|}, \\
 \Gamma_z = \frac{1}{2}\sum \limits_\sigma\frac{\sigma t^2_{\sigma}}{|v_{\sigma,z}|}, \\
  h_{eff}=\frac{1}{8\tau_F}\sum \limits_\sigma \frac{\sigma t^2_{\sigma}}{|v_{\sigma,z}|\varepsilon_{F,\sigma,z}}
  \enspace ,
  \label{eilenberger_parameters_3D}
  \end{eqnarray}

  where $\varepsilon_{F,\sigma,z}=p_{\sigma,z}^2/2m_F$ and $p_{\sigma_z}$ is defined by $(p_{\sigma,z}^2+p_S^2)/2m_F=\mu + \sigma h$. As usual, we assume that $\tau_F^{-1} \ll \varepsilon_{F,\sigma,z}$ and Eq.~(\ref{Green_bulk_ferr_3D}) is expanded to the first order with respect to the parameter $(\tau_F\varepsilon_{F,\sigma,z})^{-1}$.

  In the considered case $(\tau_F\varepsilon_{F,\sigma,z})^{-1} \ll 1$ the effective exchange field $h_{eff}$ described by Eq.~(\ref{eilenberger_parameters_3D}) is much smaller than $\Gamma$ and $\Gamma_z$, nevertheless we keep it in the equation, because it is the only source of the singlet-triplet conversion. The other source of $h_{eff}$ is the spin-mixing upon reflecting the S/F interface \cite{Cottet2009,Eschrig2015}. It is not accounted for in the framework of our simple model, but it can be taken into account making use of the general boundary conditions to the Usadel equation, as it is demonstrated in the main text.

  The effective parameters for the case of 2D ferromagnet take the form:
  \begin{eqnarray}
  \Gamma = \frac{1}{4\tau_F}\sum \limits_\sigma t_\sigma^2 \frac{1}{(-\xi_F+\sigma h)^2 + 1/(4\tau_F^2)}, \\
  \Gamma_z = \frac{1}{4\tau_F}\sum \limits_\sigma \sigma t_\sigma^2 \frac{1}{(-\xi_F+\sigma h)^2 + 1/(4\tau_F^2)}, \\
  h_{eff}=\frac{1}{2}\sum \limits_\sigma \sigma t_\sigma^2 \frac{\xi_F - \sigma h}{(-\xi_F+\sigma h)^2 + 1/(4\tau_F^2)},
  \enspace ,
  \label{eilenberger_parameters_2D}
  \end{eqnarray}

where $\xi_F = p_S^2/2m_F - \mu$. It is seen that for the case of extremely thin ferromagnet the impurity scattering is not important for generation of the effective exchange. On the contrary, the effective deparing parameters $\Gamma$ and $\Gamma_z$ are only nonzero due to the presence of impurities in the ferromagnet because there is no leakage of the correlations into the depth of the ferromagnet in this case.

In the Matsubara representation Eq.~(\ref{eilenberger_final}) takes the form:
\begin{eqnarray}
  i \bm v_S \partial_{\bm r}\check g_{S,l}(\bm r, \bm p_S)+\Bigl[i (\omega+{\rm sgn} \omega \Gamma) \hat\tau_z  + i{\rm sgn} \omega \Gamma_z \hat\sigma_z - \nonumber \\
 \check \Sigma_S - M_{kj}^S \hat\sigma_k p_{S,j} - \check \Delta (\bm r)+h_{eff} \hat\sigma_z \hat\tau_z,\check g_{S,l} \Bigr] = 0
  \enspace .~~~~~~~
  \label{eilenberger_final_m}
  \end{eqnarray}
 
 {\it Generalized Usadel equation.}

 As usual, in the dirty case we seek for the solution of the Eilenberger equation in the form $\check g_{S,l}(\bm r, \bm p_S)=\check g(\bm r)+\check {\bm g}_a(\bm r)\bm n$ with $\bm n = \bm p_S/p_S$ and $\check g_a(\bm r) \ll \check g(\bm r)$. Averaging Eq.~(\ref{eilenberger_final_m}) over the 2D superconducting Fermi-surface we obtain:
 \begin{eqnarray}
 \frac{i  v_S}{2}\partial_{\bm r} \check {\bm g}_a+ \Bigl[ i (\omega  +  {\rm sgn}\omega \Gamma) \hat\tau_z + h_{eff} \hat\tau_z \hat\sigma_z + i {\rm sgn}\omega \Gamma_z \hat\sigma_z - \nonumber \\
 \check \Delta (\bm r), \check g \Bigr]-\Bigl[\frac{1}{2}\bm M_k^S \hat\sigma_k p_S , \check {\bm g}_a \Bigr]=0,~~~~~~
\label{usadel_1}
\end{eqnarray}
where $\bm M_k^S = (M_{kx}^S, M_{ky}^S, M_{kz}^S)$. Further we multiply Eq.~(\ref{eilenberger_final}) by $\bm n$ and average over the Fermi-surface. Taking into account that $\check \Sigma_S=-(i/2\tau_S)\check g(\bm r)$ and that $\check g_a \check g = - \check g \check g_a$, following from the normalization condition, we obtain:
\begin{eqnarray}
\check {\bm g}_a =  -v_S \tau_F \check g \partial_{\bm r} \check g + \frac{\tau_F}{i} \check g \Bigl[ \bm M_k^S \hat\sigma_k p_S , \check g \Bigr].
\label{usadel_2}
\end{eqnarray}
Introducing operator $\hat \partial_{\bm r}=\partial_{\bm r} + i \bigl[ \bm M_k^S \hat\sigma_k m_S, ... \bigr]$, we come to the following Usadel equation:
\begin{eqnarray}
- D \hat \partial_{\bm r}\bigl( \check g \hat \partial_{\bm r} \check g \bigr) + \Bigl[ \omega \hat\tau_z +  {\rm sgn} \omega \Gamma \hat\tau_z  -i h_{eff} \hat\tau_z \hat\sigma_z +  \nonumber \\
 {\rm sgn} \omega \Gamma_z \hat\sigma_z +i\check \Delta (\bm r), \check g \Bigr]=0.~~~~
\label{usadel_final_matsubara}
\end{eqnarray}
Denoting $P=\Gamma_z/\Gamma$ we see that it coincides with Eq.~(\ref{usadel_local}) in the case where $\bm P(\bm r)$ is always aligned with $h_{eff}(\bm r)$. This assumption arises naturally in our model of strong ferromagnet.
 
It is also worth noting that the value of the effective spin-orbit coupling generated by the magnetization texture can be rather large and is controlled by the characteristic scale of the magnetic inhomogeneity. It is restricted by the assumption of slow variation of the magnetization, but, nevertheless, it is not proportional to the tunnel probability $t^2$. Together with the nonzero $\Gamma_z$ term this effective spin-orbit coupling generates different configurations of the spontaneous electric currents or inhomogeneous superconducting states in the superconducting film.


\bibliography{MVbibME}
\end{document}